\documentclass[twocolumn,showpacs,preprintnumbers,amsmath,amssymb]{revtex4}
\usepackage{graphicx}% Include figure files
\usepackage{dcolumn}% Align table columns on decimal point
\usepackage{bm}% bold math

\begin{document}

\preprint{APS/123-QED}

\title{Gapped ground state in a spin-1/2 frustrated square lattice}

\author{H. Yamaguchi$^{1}$, N. Uemoto$^{1}$, T. Okubo$^{2}$, Y. Kono$^{3}$, S. Kittaka$^{3}$, T. Sakakibara$^{4}$, T. Yajima$^{4}$, S. Shimono$^{5}$, Y. Iwasaki$^{1}$, and Y. Hosokoshi$^{1}$}
% \altaffiliation[Also at ]{Physics Department, XYZ University.}%Lines break automatically or can be forced with \\
%\author{Second Author}
%\email{yamaguchi@p.s.osakafu-u.ac.jp}
\affiliation{
$^1$Department of Physical Science, Osaka Prefecture University, Osaka 599-8531, Japan\\
$^2$Department of Physics, the University of Tokyo, Tokyo 113-0033, Japan\\
$^3$Department of Physics, Chuo University, Tokyo 112-8551, Japan\\ 
$^4$Institute for Solid State Physics, the University of Tokyo, Chiba 277-8581, Japan\\
$^5$Department of Materials Science and Engineering, National Defense Academy, Kanagawa 239-8686, Japan\\
}

%\author{Charlie Author}
%\homepage{http://www.Second.institution.edu/~Charlie.Author}
%\affiliation{
Second institution and/or address\\
This line break forced% with \\

\date{\today}% It is always \today, today,
             %  but any date may be explicitly specified

\begin{abstract}
We present a model compound with a spin-1/2 frustrated square lattice, in which three ferromagnetic (F) interactions and one antiferromagnetic (AF) compet. 
Considering the effective spin-1 formed by the dominant F dimer, this square lattice can be mapped to a spin-1 spatially anisotropic triangular lattice.  
The magnetization curve exhibits gapped behavior indicative of a dominant one-dimensional (1D) AF correlation.
In the field-induced gapless phase, the specific heat and magnetic susceptibility show a phase transition to an ordered state with two-dimensional (2D) characteristics.
These results indicate that the spin-1 Haldane state is extended to the 2D system.
We demonstrate that the gapped ground state observed in the present spin-1/2 frustrated square lattice originates from the one-dimensionalization caused by frustration.
\end{abstract}

% PACS, the Physics and Astronomy
                             % Classification Scheme.
%\keywords{Suggested keywords}%Use showkeys class option if keyword
                              %display desired

\maketitle
%frustrated square latticeから始める、最近の実現例
%強磁性相関を１つ含む場合が本系　⇒　その場合は実効的にS=1三角格子となる
Quantum phenomena associated with magnetic frustration, in which neighboring spins interact by competing exchange interactions that cannot be satisfied simultaneously, have been a key focus of research in condensed matter physics.
Triangular lattice antiferromagnets are archetypal examples of frustrated magnets.
The quantum spin liquid picture on the spin-1/2 triangular lattice proposed in 1973 has stimulated extensive research on quantum ground states caused by frustration~\cite{RVB}. 
Experimental efforts to establish this exotic quantum phase in triangular-based systems are still actively ongoing~\cite{frust_rev1,frust_rev2}.
%However, subsequent numerical studies have established that its ground state is an antiferromagnetic (AF) order~\cite{120_1,120_2}.
For square lattice, there is no geometrical frustration; however, frustration can be caused by the coexistence of ferromagnetic (F) and antiferromagnetic (AF) interactions. 
Such a system has been theoretically studied in association with Josephson-junction arrays~\cite{FFSL_1,FFSL_2,FFSL_3}, but there have been no reports of ideal model compounds. 
Recently, our spin arrangement design using organic radicals enabled the formation of spin-1/2 frustrated square lattices.
These compounds demonstrate that strong quantum fluctuations and frustration effects stabilize quantum states~\cite{TCNQ, PF6, SbF6}.
The spin model focused on in this study is one example of such a spin-1/2 frustrated square lattice, where three F interactions and one AF interaction form a unit square, as shown in Fig. 1(a).
Furthermore, because the ferromagnetically coupled spins can form an effective spin-1 state, this system can be mapped to a spin-1 spatially anisotropic triangular lattice, as shown in Fig. 1(b).

%歪んだ三角格子の理論研究、相図
%1/2の系でone-dimensionality by frustrationと多様な相図を紹介　⇒　S=1の系ではgappedなHaldaneが基底である特異性、そしてHaldaneと秩序の間に臨界点があり予想されているとこと
Frustrated systems have a delicate energy balance, and thus, additional factors can easily change their ground states. 
One such factor is spatial anisotropy, which is equivalent to lattice distortion.
The spin-1/2 Heisenberg triangular lattice case has been extensively studied.
The spatial anisotropy between decoupled one-dimensional (1D) chain and isotropic triangular lattice, where 1D chains are coupled through frustrated zigzag paths, induces a dimensional reduction owing to the frustrated interchain couplings~\cite{Cs2CuCl4_exp1, Cs2CuCl4_exp2, spin1/2_1, spin1/2_2, spin1/2_3,spin1/2_4}.
As a result, a gapless disordered state analogous to a Tomonaga-Luttinger liquid (TLL) in a 1D system becomes stable in the spin-1/2 anisotropic triangular lattice, which is the so-called one-dimensionalization.
For the spin-1 case, which corresponds to the mapped model in the present work, a dimensional reduction associated with the spatial anisotropy has also been suggested~\cite{spin1_5,spin1_7, spin1_6,spin1_1}.
Cruciallly, unlike the gapless ground state in the spin-1/2 chain, a gapped Haldane state is formed in the spin-1 chain~\cite{haldane}.
Examining whether the one-dimensionalization caused by frustration stabilizes the gapped Haldane state is an important experimental issue.

%A magnetic ordered phase appears when the interchain frustrated zigzag couplings close the Haldane gap.
%The one-dimensionalization caused by the frustration is predicted to enhance the critical coupling required to close the Haldane gap by an order of magnitude compared to unfrustrated case~\cite{spin1_5,spin1_7, spin1_6, spin1_1, spin1_square1, spin1_square2}. 

%, where 1D chains are formed through $J$ and coupled through frustrating zigzag path $J^{'}$, 
%On the other hand, for the spin-1 case has been little explored in the literature. 
%the Haldane gap is only suppressed at J/J ≈ 0.4 [19], ascompared with J/J ≈ 0.04 for the unfrustrated case [20]. 

%要約
In this letter, we present a model compound with a spin-1/2 frustrated square lattice.
We successfully synthesized single crystals of the verdazyl-based complex $[$Zn(hfac)$_2]$(4-Br-$o$-Py-V).
Molecular orbital (MO) calculations indicate the formation of a square lattice composed of three F interactions and one AF interaction causing frustration, which can be mapped to the spin-1 spatially anisotropic triangular lattice.
The magnetization curve exhibits gapped behavior, indicative of a dominant 1D AF correlation.
In the field-induced gapless phase, the specific heat and magnetic susceptibility show a phase transition to an ordered state with two-dimensional (2D) characteristics.
These results demonstrate the realization of the spin-1 Haldane state in the spatially anisotropic triangular lattice owing to the one-dimensionalization caused by frustration.

%実験方法
We prepared 4-Br-$o$-Py-V through a conventional procedure for the verdazyl radical~\cite{gosei} and synthesized $[$Zn(hfac)$_2]$(4-Br-$o$-Py-V) using a reported procedure for similar verdazyl-based complexes~\cite{Zn, Zn_uemoto, Zn_kono}.
Recrystallization using acetonitrile yielded a dark-green crystal of $[$Zn(hfac)$_2]$(4-Br-$o$-Py-V).
The crystal structure was determined from the intensity data obtained using a Bruker D8 VENTURE with a PHOTON II detector and a Rigaku AFC-8R Mercury CCD RA-Micro7 diffractometer with a Japan Thermal Engineering XR-HR10K at room temperature and 25 K. 
The magnetization was measured using a commercial SQUID magnetometer (MPMS-XL, Quantum Design) above 1.8 K and a capacitive Faraday magnetometer in a dilution refrigerator down to approximately 80 mK. 
The magnetic susceptibility was corrected for the diamagnetic contribution of $-3.5{\times}10^{-4}$ emu mol$^{-1}$, which is determined to become almost ${\chi}^{-1}$ $\propto$ $T$ above approximately 50 K and close to that calculated by Pascal's method.
The specific heat was measured using a handmade apparatus by a standard adiabatic heat-pulse method by using a $^3$He refrigerator down to approximately 0.3 K.
We subtracted nuclear Schottky contributions caused by the Zeeman splitting of nuclear spins, which were evaluated to be 2.43${\times}$ $10^{-4}$$H^2$/$T^2$~\cite{Nheat}.
Considering the isotropic nature of organic radical systems, all experiments were performed using small, randomly oriented single crystals.

%結晶構造
Figure 1(c) shows the molecular structure of $[$Zn(hfac)$_2]$(4-Br-$o$-Py-V), which has an $S$ = 1/2 within a molecule.
Because the focus of this study was low-temperature magnetic properties, we considered low-temperature crystallographic data at 25 K.
No indication of a structural phase transition is observed down to 25 K~\cite{supple1}.
The crystallographic parameters at 25 K are as follows: monoclinic, space group $P2_{1}/c$, $a$ = 8.784(4) $\rm{\AA}$, $b$ = 31.022(13) $\rm{\AA}$, $c$ = 11.641(6) $\rm{\AA}$, $\beta$ = 94.525(8) $^{\circ}$, $V$ = 3162(3) $\rm{\AA}^3$, and $Z$ = 4. 
We evaluated the intermolecular exchange interactions through $ab$ $initio$ MO calculations and found four types of predominant interactions~\cite{supple1, MOcal}, as shown in Fig. 1(d).
They were evaluated as $J_{\rm{F1}}/k_{\rm{B}}$ = $-$28.5 K, $J_{\rm{F2}}/k_{\rm{B}}$ = $-$3.8 K, $J_{\rm{F3}}/k_{\rm{B}}$ = $-1.5$ K, and $J_{\rm{AF}}/k_{\rm{B}}$ = $1.5$ K, which are defined in the Heisenberg spin Hamiltonian given by $\mathcal {H} = J_{n}{\sum^{}_{<i,j>}}\textbf{{\textit S}}_{i}{\cdot}\textbf{{\textit S}}_{j}$, where $\sum_{<i,j>}$ denotes the sum over the neighboring spin pairs.
These interactions lead to the formation of a spin-1/2 square lattice in the $ac$ plane, as shown in Fig. 1(a).
In our previous studies, we confirmed that the MO calculations for verdazyl-based compounds provide reliable values of exchange interactions to qualitatively examine their intrinsic behavior\cite{Zn, Zn_uemoto, Zn_kono, 3Cl4FV, 2Cl6FV, pBrV, b26Cl2V, a235Cl3V}.
The nonmagnetic Zn(hfac)$_2$ acts as a spacer between the 2D structures; hence, the 2D characteristics of the present spin model are enhanced.
%Considering strong dependence on the calculation method and basis set in the MO calculation, such small interchain couplings do not have enough reliability~\cite{MOseido}.
Furthermore, two spins coupled by the strongest ferromagnetic interaction $J_{\rm{F1}}$ can be considered an effective spin-1 in the low-temperature region, forming a spin-1 spatially anisotropic triangular lattice, as shown in Fig. 1(b).

 %磁化率 T>2 K、磁化曲線
Figure 2(a) shows the temperature dependence of the magnetic susceptibility ( $\chi$ = $M/H$) at 0.5 T. 
A broad peak appears at approximately 3 K, which indicates an AF short-range order.
The temperature dependence of $\chi T$ increases with decreasing temperature till approximately 20 K, which indicates the dominant contributions of the F exchange interactions. 
We fitted $\chi$ using the Curie-Weiss law for $T$ $\textgreater$ 50, and the Weiss temperature was estimated to be $\theta_{\rm{W}}$ = $2.7$(2) K. 
Thus, the mean-field approximation with $\theta_{\rm{W}}$ indicated ($J_{\rm{F1}}$+$J_{\rm{F2}}$+$J_{\rm{F3}}$+$J_{\rm{AF}}$)/$k_{\rm{B}}$ $\sim$ $-10.8$ K.
Comparing this value with the MO evaluations shows that the F and AF interactions evaluated from the MO calculations are overestimated and underestimated, respectively.
Figure 2(b) shows the magnetization curve at 85 mK. 
A saturation value of 0.98 $\mu_{B}/f.u.$ indicates that the purity of the radicals was approximately 98 $\%$.
In the low-field region, the slope is particularly small, which is reminiscent of gapped behavior.
The field derivative of the magnetization curve ($dM/dH$) clearly exhibits a discontinuous change at approximately 0.3 T, indicative of the critical field where the spin gap closes.  
Furthermore, we evaluated the phase transition field to the fully-polarized phase from the midpoint of the drastic change in $dM/dH$.

%The upturn below approximately 10 K is caused by slight paramagnetic impurities associated with lattice defects, which is commonly observed in verdazyl-based compounds~\cite{PRL,Zn, TCNQ_JPSJ}.
%Assuming conventional paramagnetic behavior $C_{\rm{imp}}/T$ and gapless excitations, where $C_{\rm{imp}}$ is the Curie constant of the impurities, the paramagnetic impurities is evaluated to be approximately 2.7 $\%$ of all spins, which is close to that evaluated in the verdazyl-based salt with similar molecular structure~\cite{3Dhoneycomb}.

%磁化率 低温
Figure 3(a) and 3(b) show the low-temperature behavior of $\chi$ and its temperature derivative $d{\chi}/dT$, respectively.
We observed an extremum in each $\chi$, which switched from minimum to maximum with increasing field.
These results are similar to those of the crossover temperature to a TLL regime in 1D gapped spin systems~\cite{the1,the2,the3, BP_CM, DY_M, 3Br4FV}. 
At lower temperatures, a sharp peak was observed for $d{\chi}/dT$.
These peaks indicate a phase transition to a 3D AF order, and the transition temperatures are consistent with the following specific heat results.

%比熱
The temperature dependence of the specific heat $C$ is shown in Fig. 4(a). 
In the low-temperature region discussed here, the magnetic contributions are expected to be dominant, as observed in other verdazyl-based compounds~\cite{Zn, pBrV, b26Cl2V, a235Cl3V}. 
There is no peak showing a phase transition at zero field, and sharp peaks associated with the phase transition to the ordered state appear for 1.0 T $\textless$ $H$ $\textless$ 4.5 T.
These phase transition temperatures were plotted in a field-temperature phase diagram, as shown in Fig. 4(b).
The obtained phase boundary is consistent with the peak temperatures of $d{\chi}/dT$ and the transition fields of $dM/dH$.
Note that the domelike shape of the phase boundary is typical of gapped quantum spin systems. 
In the ordered phase, $C/T^2$ appoaches a constant value at sufficiently low temperatures, showing $C$ $\propto$ $T^2$, as shown in Fig. 4(c). 
This $T^2$ dependence of the low-temperature specific heat indicates gapless linear magnon dispersions in the 2D AF system, confirming that the AF ordered state is based on the expected square lattice.
Figure 4(d) shows the entropy obtained via the integration of $C/T$. 
Considering that the entropy change associated with the phase transition is close to 1/3 of the total magnetic entropy of 5.76 ($R$ln2), the present spin system indeed has a 2D character with a sufficient development of short-range order above $T_{\rm{N}}$.

%考察 
%秩序相以外は１D的でギャップやTLLの存在
%⇒　一方で、秩序では２D的長距離秩序　⇒　１D性はフラストレーションによって無理矢理
We considered the ground state of the present spin-1/2 square lattice.
The gapped behavior of the magnetization curve and the domelike phase boundary demonstrate effective interactions forming a spin gap.
Furthermore, the broad extremum of the magnetic susceptibility, which is interpreted as the crossover to the TLL region~\cite{the1, the2, the3}, appeared at temperatures slightly above the phase boundary.
Such a phase diagram is characteristic of weakly coupled 1D gapped systems~\cite{BP_CM, DY_M, 3Br4FV, 3Br4FV_kono}. 
Assuming the effective spin-1 formed by the strong F interaction $J_{\rm{F1}}$, we considered the magnetic state in the mapped spin-1 triangular lattice. 
Because the observed 1D AF behavior has its origin in the only AF interaction $J_{\rm{AF}}$, the gapped ground state should be associated with the spin-1 Haldane state formed by $J_{\rm{AF}}$, as shown in Fig. 1(b). 
Theoretical studies on the spin-1 spatially anisotropic triangular indicate that one-dimensionalization caused by frustration can stabilize a Haldane state extended in a 2D system~\cite{spin1_5,spin1_7, spin1_6,spin1_1}.
%Although the present mapped triangular lattice differs in that the interchain couplings are F, the frustrated interactions between 1D chains are also expected to stabilize such an extended Haldane state.
Although the present mapped triangular lattice differs in that the interchain couplings are F, such an extended Haldane state is expected to be stabilized. 
As a toy model of a coupled Haldane chain, suppose we connect two $S$ = 1/2 dimers through two Heisenberg interactions, forming a square. 
One can easily show that the ground state of the model with the F interdimer interactions is closer to the decoupled singlets than the AF interactions with the same amplitude. 
Therefore, the F interchain interactions have less effect on a decoupled Haldane state than AF interchain interactions.
This fact also supports the stabilization of the extended Haldane state in the present 2D system.

In the field-induced gapless phase, a spiral magnetic state is expected to be realized, and the low-temperature specific heat exhibits contributions associated with the magnon dispersion in a 2D AF system.
Thus, the gradual increase in the magnetization curve observed in the present system can also be interpreted as a characteristic of the 2D AF system.
If we assume a decoupled spin-1 Haldane chain with a spin gap of $\sim$0.3 T, the magnetization curve exhibits a more drastic increase with a sharp double-peak structure of $dM/dH$ and saturates at $\sim$3.5 T, which is clearly different from the observed behavior.
These results indicate that the actual interactions in the present system form the expected 2D system.
Hence, the observed 1D quantum ground state should be stabilized by dimensional reduction caused by frustration.

%結論
In summary, we successfully synthesized single crystals of the verdazyl-based complex $[$Zn(hfac)$_2]$(4-Br-$o$-Py-V). 
The MO calculations indicated that three F interactions and one AF interaction form a spin-1/2 frustrated square lattice. 
Considering the effective spin-1 formed by the most dominant F interaction, this square lattice can be mapped to a spin-1 spatially anisotropic triangular lattice.  
It should be noted that the magnetization curve exhibited gapped behavior.
In the field-induced gapless phase, the magnetic susceptibility indicated an extremum associated with the crossover to the TLL regime.
At the lower temperatures, the specific heat and magnetic susceptibility indicated temperature dependencies associated with a phase transition to an AF ordered state.
The obtained domelike phase boundary and the TLL phase slightly above it are typical of 1D gapped spin systems.
In addition to these 1D characteristics, the $T^2$ dependence in the low-temperature specific heat and a gradual increase in the magnetization curve show 2D characteristics in the ordered phase.
These results demonstrate the realization of the spin-1 Haldane state extended to a 2D system. 
Accordingly, the gapped ground state observed in the present spin-1/2 square lattice originates from the one-dimensionalization caused by frustration.
The present material provides a 2D quantum system with a gapped ground state stabilized by frustrated interactions and will stimulate further studies on quantum phenomena associated with square-based frustration.

\begin{figure}[t]
\begin{center}
\includegraphics[width=20pc]{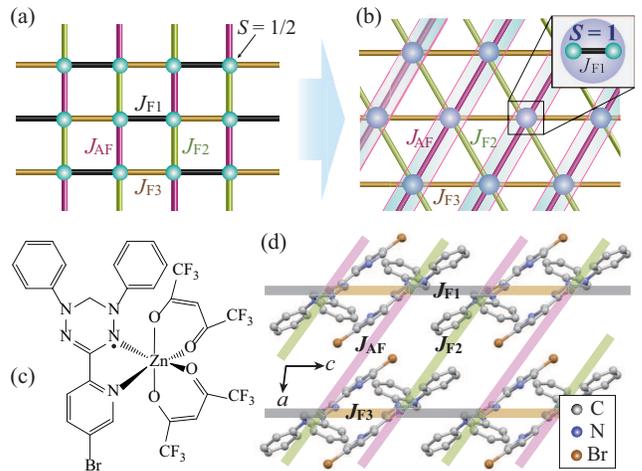}
\caption{(color online) (a) Spin-1/2 frustrated square lattice composed of F interactions, $J_{\rm{F1}}$, $J_{\rm{F2}}$, and $J_{\rm{F3}}$, and AF interaction $J_{\rm{AF}}$ in $[$Zn(hfac)$_2]$(4-Br-$o$-Py-V). 
(b) Spin-1 spatially anisotropic triangular lattice in terms of the effective spin-1 formed through $J_{\rm{F1}}$. Thick colored lines along the $J_{\rm{AF}}$ chain represent the Haldane state with a spin gap. 
(c) Molecular structure of $[$Zn(hfac)$_2]$(4-Br-$o$-Py-V). (d) Crystal structure forming the square lattice in the $ac$ plane. Hydrogen atoms and Zn(hfac)$_2$ are omitted for clarity.}
\end{center}
\end{figure}

\begin{figure}[t]
\begin{center}
\includegraphics[width=20pc]{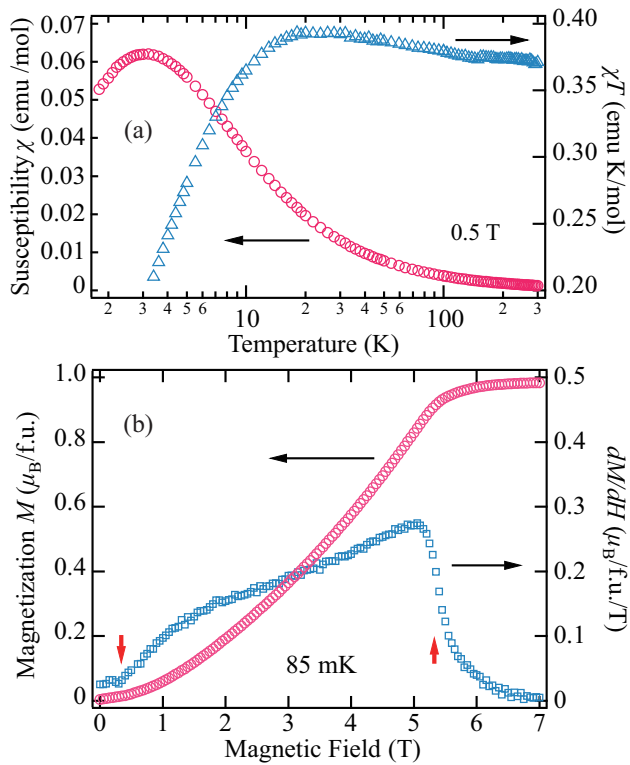}
\caption{(color online) (a) Temperature dependence of the magnetic susceptibility ($\chi$ = $M/H$) and $\chi$$T$ of $[$Zn(hfac)$_2]$(4-Br-$o$-Py-V) at 0.5 T . 
(b) Magnetization curve and its field derivative of $[$Zn(hfac)$_2]$(4-Br-$o$-Py-V) at 85 mK. Thick arrows indicate the phase transition fields.}\label{f2}
\end{center}
\end{figure}

\begin{figure}[t]
\begin{center}
\includegraphics[width=20pc]{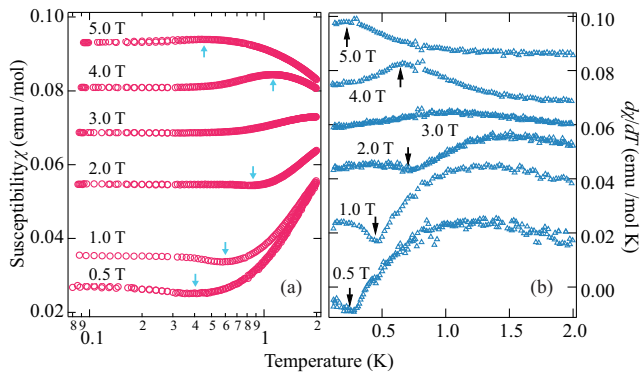}
\caption{(color online) Low-temperature region of (a) the magnetic susceptibility ($\chi$ = $M/H$) and (b) its temperature derivative $d\chi/dT$ of $[$Zn(hfac)$_2]$(4-Br-$o$-Py-V) in various magnetic fields. Arrows indicate the broad extremum of $\chi$ and phase transition peak of $d\chi/dT$.
}\label{f3}
\end{center}
\end{figure}

\begin{figure}[t]
\begin{center}
\includegraphics[width=20pc]{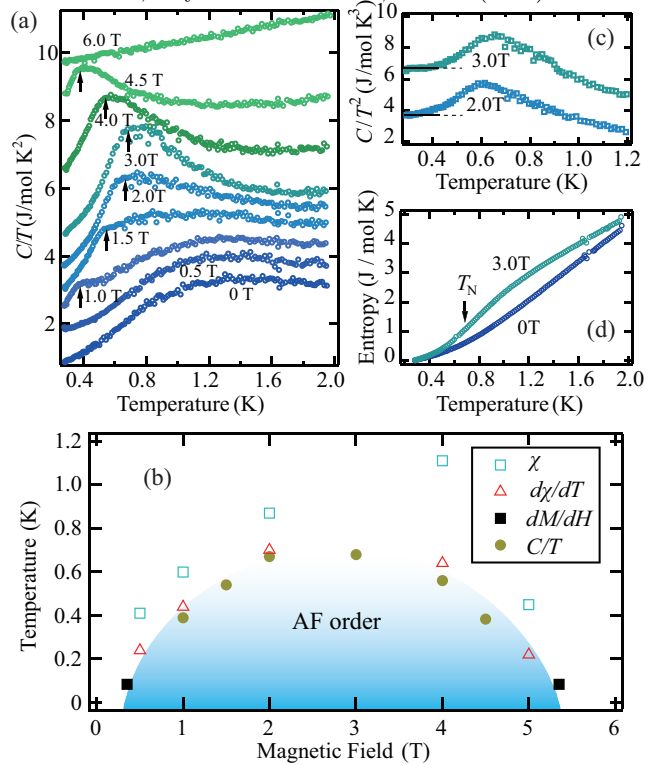}
\caption{(color online) (a) Temperature dependence of the specific heat $C/T$ of $[$Zn(hfac)$_2]$(4-Br-$o$-Py-V). 
For clarity, the values for 0.5, 1.0, 1.5, 2.0, 3.0, 4.0, 4.5, and 6.0 T have been shifted up by 0.54, 1.3, 2.0, 2.7, 3.5, 4.9, 6.3, and 8.5 J/ mol K$^2$, respectively.
(b) Magnetic field vs. temperature phase diagram, showing the domelike phase boundary of the 3D AF order. Open squares obtained from the extremum temperature of $\chi$ are related to the crossover to the TLL.
(c) Low-temperature region of $C/T^2$ at 2.0 and 3.0 T. Solide lines are guides that show the asymptotic behavior to a constant value. For clarity, the values for 3.0 has been shifted up by 2.5 J/ mol K$^3$. 
(d) Entropy obtained through integration of $C/T$ at 0 and 3.0 T.
}\label{f4}
\end{center}
\end{figure}

\begin{acknowledgments}
This research was partly supported by Grant for Basic Science Research Projects from KAKENHI (No. 19J01004, No. 19H04550, and 19K03740), JST, PRESTO Grant Number JPMJPR1912, the Murata Science Foundation, Kyoto Technoscience Center, SEI Group CSR Foundation, and Nanotechnology Platform Program (Molecule and Material Synthesis) of the Ministry of Education, Culture, Sports, Science and Technology (MEXT), Japan. A part of this work was performed as the joint-research program of ISSP, the University of Tokyo and the Institute for Molecular Science.
\end{acknowledgments}

%%%%%%%%%%%%%%%%%%%%%%%%%%%%%%%%%%%%%%%%%%%%%%%%%%%%%%%%%%%%%%
%%%%%%%%%%%%


\begin{thebibliography}{99}
\bibitem{MOcal}
The MO calculations were performed using the UB3LYP method in the Gaussian 09 program package. 
The basis set is 6-31G. 
For the estimation of intermolecular magnetic interaction, we applied the previously reported evaluation scheme~\cite{MO_method}.
\end{thebibliography}
\end{document}